\documentclass{article}
\usepackage{amsmath,amssymb,amsthm,amsfonts}
\usepackage{bbm}
\usepackage{graphicx,epsfig,color}
\usepackage[margin=3cm]{geometry}

\numberwithin{equation}{section}

\newcommand{\id}{{\boldsymbol{\mathbbm{1}}}}
\DeclareMathOperator{\tr}{tr}
\DeclareMathOperator{\sym}{sym}
\DeclareMathOperator{\dev}{dev}
\DeclareMathOperator{\skewp}{skew}
\DeclareMathOperator{\curl}{curl}
\DeclareMathOperator{\Curl}{Curl}
\DeclareMathOperator{\grad}{grad}
\DeclareMathOperator{\divv}{div}
\DeclareMathOperator{\Divv}{Div}
\DeclareMathOperator{\polar}{polar}

\begin{document}

\title{Soliton-like solutions based on geometrically nonlinear Cosserat micropolar elasticity}
\author{
  Christian G. B\"ohmer\footnote{Corresponding author: C. G. B\"ohmer, Department of Mathematics, University College London, Gower Street, London, WC1E 6BT, UK, email: c.boehmer@ucl.ac.uk}
  \ and
  Patrizio Neff\footnote{P. Neff, Fakult\"at f\"ur Mathematik, Universit\"at Duisburg-Essen, Thea-Leymann-Stra\ss e 9, 45127 Essen, Germany, email: patrizio.neff@uni-due.de} 
  \ and 
  Belgin Seymeno\u{g}lu\footnote{B. Seymeno\u{g}lu, Department of Mathematics, University College London, Gower Street, London, WC1E 6BT, UK, email: belgin.seymenoglu.10@ucl.ac.uk}
}
\date{}
\maketitle

\begin{abstract}
The Cosserat model generalises an elastic material taking into account the possible microstructure of the elements of the material continuum. In particular, within the Cosserat model the structured material point is rigid and can only experience microrotation, which is also known as micropolar elasticity. We present the geometrically nonlinear theory taking into account all possible interaction terms between the elastic and microelastic structure. This is achieved by considering the irreducible pieces of the deformation gradient and of the dislocation curvature tensor. In addition we also consider the so-called Cosserat coupling term. In this setting we seek soliton type solutions assuming small elastic displacements, however, we allow the material points to experience full rotations which are not assumed to be small. By choosing a particular ansatz we are able to reduce the system of equations to a Sine-Gordon type equation which is known to have soliton solutions. 
\end{abstract}

\mbox{}\\ 
\mbox{}


\mbox{}


\mbox{}

\section{Introduction}

The theory of classical elasticity is based on the assumption that the material points are structureless and their possible internal properties like an orientation are disregarded. The Cosserat~\cite{Cosserat09} model is a generalisation which takes into account a possible microstructure of the elements of the material continuum. In the Cosserat continuum material points can experience rotations without stretches. Therefore, in addition to the usual deformation field $\varphi$ there is an independent rotation field $\overline{R}$. This idea motivated a rich variety of models with sometimes differing names~\cite{Ericksen:1957,Toupin:1962,Ericksen:1962a,Ericksen:1962b,Mindlin1964,Toupin:1964,Eringen:1964a,Eringen:1964b,Green:1964,Ericksen:1967,Schaefer:1967,Eringen:1999}.

Let us consider a material point and attach to it an arbitrary 3-frame. This means introducing nine new degrees of freedom which are 3 rotations, 1 volume expansion and 5 shear deformations. This general continuum is known as the micromorphic continuum~\cite{Eringen:1999,Neff_Forest_jel05}. One can now formulate the special case where the structured material point is rigid  and therefore it can only experience microrotation, this is the subject of Cosserat elasticity or micropolar elasticity which now has 6 degrees of freedom in total, three deformations $\varphi$ and three microrotations $\overline{R}$, considered in the present work. 

This theory has two limits: In the case of no rotations one recovers standard elasticity theory, however, the limit of no deformations is also very interesting. Models of this type can in fact be traced back to MacCullagh, see~\cite{Whittaker_VolI}, who was interested in understanding the ether. Since rotations in 3 dimensions do not commute in general, the resulting theory of microrotations is generically always nonlinear.

The Cosserat model has been theoretically proposed by the Cosserat brothers in a completely nonlinear setting \cite{Cosserat09}. However, they never proposed any specific constitutive relation. Their work was largely forgotten for nearly half a century when interest arose again in the late 1950s. Then, at first, scientists looked at a constraint model, in which the Cosserat microrotations are assumed to coincide with the continuum rotation. The investigations were mainly carried out in a linearized framework, see the recent acount given in \cite{Neff_Ghiba_unified_micromorph13}. Later, the full concept of free microrotations was rediscovered and soon extended along the given lines to micromorphic models, see \cite{Neff_Ghiba_unified_micromorph13} for a summary. It was quickly realized that these extended continuum mechanical models give rise to wave dispersion. Some intriguing new facets of linear micromorphic models being able to describe complete band gaps have only be discovered recently \cite{Neff_Ghiba_band_gaps_micromorph13,Neff_Ghiba_band_gaps_micromorph_zamm14}. Interestingly, this new feature really needs the micromorphic kinematics, stricly extending the Cosserat approach. The well-posedness for the static and dynamic case in the linearized setting is well-established. The static problem in a variational framework is also partially understood in the finite strain case, for both the Cosserat and the micromorphic model \cite{Neffexistence:2014,Neff_curl06,Neff_Cosserat_plasticity05}.

The geometrically nonlinear micromorphic model was studied in~\cite{Neff_micromorphic_rse_05,Neff_Forest_jel05,Neff_Muench_simple_shear09} where first existence theorems could be proved. To our knowledge, there are no known existence theorems for geometrically nonlinear Cosserat or micromorphic dynamic models. This point notwithstanding, in this paper we consider the geometrically nonlinear Cosserat model with a nonstandard constitutive coupling between Cosserat rotations and the deformation gradient. There are already many works taking nonlinear models with microstructure as a starting point for soliton-like solutions. However, most works seem to begin with a one-dimensional framework right away, while in our model we consider the fully 3-dimensional setting in the first place, see also~\cite{Maugin_Miled:1986a,Maugin_Miled:1986b,Puget_Maugin:1989,Kivshar_Malomed:1990,Sayadi_Puget:1991,Zorski_Infeld:1992,Potapov:1999}. It is only when we introduce a special ansatz that the Cosserat model simplifies. The simplification we use consists in assuming small displacements but allowing for arbitrarily large rotations. In doing so we keep the geometrically exact curvature expression and consider a certain nonlinear coupling between displacements and rotations.

On the other hand, in~\cite{boehmer2011rota,boehmer2012gauge,boehmer2013rota} exact wave-like solutions were found in the geometrically nonlinear Cosserat model. The existence of such solutions is a non-trivial fact, these solutions were first constructed using spinor methods~\cite{boehmer2011rota} which significantly simplified the Euler-Lagrange equations. This made it possible to find plane-wave solutions. Subsequent work~\cite{boehmer2012gauge,boehmer2013rota} followed a more direct approach where wave-like solutions could be constructed without resorting to spinor methods. The study of wave type solutions in continua with or without micro-structure has always been of interest. The existence of linear elastic waves is well known and has motivated subsequent research ever since.

A suitable deformation measure in Cosserat elasticity is the so-called dislocation curvature tensor $\overline{R}^T\negthickspace \Curl \overline{R}$, see \cite{Neff_curl06}. It turns out that this commonly used naming convention is in fact slightly misleading. When Cosserat elasticity is formulated in a differential geometric framework, then the orthogonal matrix $\overline{R}$ is interpreted as the frame or tetrad of a flat manifold with torsion. The object $\overline{R}^T\negthickspace \partial_i \overline{R}$ is now a rank 3 tensor which is exactly the (Weitzenb\"ock) connection of this space, or the contortion tensor. The matrix $\overline{R}^T\negthickspace \Curl \overline{R}$ is related to the Hodge dual of the contortion tensor, for more details see~\cite{boehmer2011rota,boehmer2012gauge} and also~\cite{Yavari:2012,Yavari:2013}. The axial part of this connection goes by various different names, wryness tensor~\cite{Ericksen:1957}, the Nye tensor~\cite{kroener}, and second Cosserat deformation tensor~\cite{Eringen:1999}. However, we will stick with the established notation of dislocation curvature tensor henceforth.

In the above $\Curl M$ is the matrix curl, $(\Curl M)_{ij} = \partial_k M_{il} \varepsilon_{klj}$, i.e.~$\curl$ acting on the rows of $M$. Likewise we define $\Divv M$ as the matrix divergence $\partial_{j}M_{ij} \mathbf{e}_i$, again acting on rows. We will make frequent use of the Cartan-Lie decomposition of an arbitrary matrix in 3 dimensions
\begin{align}
  X = \dev \sym M + \skewp M + \frac{1}{3}\tr(M) \id
\end{align}
where $\id$ is the identity matrix, $\sym M = (M+M^T)/2$ is the symmetric part of $M$, $\skewp M = (M-M^T)/2$ is the anti-symmetric part and $\dev M = M - 1/3 \tr(M)\id$ is the deviatoric or trace-free part of $M$.

We also exploit the fact that every orthogonal matrix can be written as the matrix exponential of a skew-symmetric matrix, namely $\overline{R} = \exp(\overline{A})$ where $\overline{A}$ is the skew-symmetric matrix generating the rotation. Furthermore, we follow the standard conventions of denoting by $\varphi$ the deformation vector, $\mathbf{u}$ is the displacement vector and $F=\nabla\varphi=\id+\nabla\mathbf{u}$ being the deformation gradient. We define $\overline{U} = \overline{R}^T F$ which is the non-symmetric right stretch tensor, also called the first Cosserat deformation tensor. In this context the second Cosserat deformation tensor is the above mentioned (Weitzenb\"ock) connection $\overline{R}^T \partial_i \overline{R}$.

We remark that we write $\overline{R}$ and $\overline{U}$ with superposed bars in order to distinguish them from the factors $R$ and $U$ of the classical polar decomposition $F = R\,U$, in which $R = \polar(F)$ is orthogonal and $U$ is positive definite, symmetric, which is the standard notation in elasticity. In particular this means $U = \sqrt{F^T F}$ and $R = \polar(F) = F (F^T F)^{-1/2}$.

\section{The complete dynamical Cosserat problem}

We write the energy functional for elastic deformation as
\begin{align}
  V_{\rm elastic}(F,\overline{R}) =
  \mu\, \| \sym (\overline{R}^T F) - \id\|^2 +
  \frac{\lambda}{2}\tr(\sym(\overline{R}^T F) - \id)^2,
  \label{Velas}
\end{align}
where $\lambda$ and $\mu$ are the standard Lam\'{e} parameters. For the dynamical treatment, we will need to subtract kinetic energy which is of the form $\rho/2 \|\dot{\varphi}\|^2$.

The energy functional of the micro-rotations $V_{\rm rotational}$ can be written in the form
\begin{align}
  V_{\rm curvature}(\overline{R})
  = {} &\kappa_1 \,\| \dev \sym (\overline{R}^T\negthickspace \Curl \overline{R})\|^2 +
  \kappa_2\, \| \skewp (\overline{R}^T\negthickspace \Curl \overline{R})\|^2
  \nonumber \\ &+
  \kappa_3 \tr(\overline{R}^T\negthickspace \Curl \overline{R})^2.
  \label{Vcurv}
\end{align}
Here $\kappa_i$ ($i=1,2,3$) are three elastic constants. As before, in order to study the dynamical problem, we will need to subtract kinetic energy which is given by $\rho_{\rm rot} \|\dot{\overline{R}}\|^2 =\rho_{\rm rot} \tr(\dot{\overline{R}}^T\dot{\overline{R}})$.

Next, we wish to introduce a coupling between the elastic displacements and the micro-rotations. In order to do so, we will `couple' the irreducible components of $\overline{R}^T\negthickspace \Curl \overline{R}$ with the corresponding irreducible parts of $\overline{R}^T F - \id$. This gives
\begin{align}
  V_{\rm interaction}(F,\overline{R})
  = {} &\chi_1 \tr(\overline{R}^T\negthickspace \Curl \overline{R})
  \tr(\overline{R}^T F)
  \nonumber \\&+
  \chi_3 \langle \dev \sym \overline{R}^T\negthickspace \Curl \overline{R},
  \dev \sym (\overline{R}^T F - \id)\rangle
  \label{Vint}
\end{align}
where $\chi_1$ and $\chi_3$ are the two coupling constants. In principle, we can also consider the skew-symmetric part of $\overline{R}^T F - \id$ in $\langle {\rm skew}\overline{R}^T\negthickspace \Curl \overline{R},
  {\rm skew} (\overline{R}^T F - \id)\rangle$. However, the term ${\rm skew}(\overline{R}^T F - \id)$ is not used for the elastic energy \eqref{Velas} and therefore we will not consider it further.

Finally, we will use the Cosserat couple term which is given by
\begin{align}
  V_{\rm coupling}(F,\overline{R}) =
  \mu_c\, \| \overline{R}^T\negthickspace \polar(F) - \id \|^2,
  \label{Vcoup}
\end{align}
where $\mu_c$ is the Cosserat couple modulus. When considering the Euler-Lagrange equations coming from this coupling term, we note that this will result in rather complicated expressions since we need the quantity $\delta \polar(F)/\delta \mathbf{u}$. When considering small displacements, the situation simplifies significantly.

\subsection{Small displacements and small rotations}

This nonlinear problem would be  very difficult to address. Therefore we consider small displacements $\|\mathbf{u}\| \ll 1$ and $\|\nabla\mathbf{u}\| \ll 1$ and small rotations $\|\overline{A}\| \ll 1$ such that $\overline{R} = \exp(\overline{A}) = \id + \overline{A} + \ldots$. Here $\overline{A}$ is a skew-symmetric matrix, $\overline{A}^T = -\overline{A}$. In the following \text{h.o.t.} will stand for higher order terms. Then
\begin{align}
  \overline{R}^T F - \id &=\exp(\overline{A})^T(\id + \nabla\mathbf{u})=
  (\id - \overline{A})(\id + \nabla\mathbf{u}) - \id + 
  \text{h.o.t.}
  \nonumber \\
  &= \id - \overline{A} + \nabla\mathbf{u} - \id + \text{h.o.t.}= \nabla\mathbf{u} - \overline{A} 
  + \text{h.o.t.} \,.
\end{align}
Therefore we have that
\begin{align}
  \tr(\overline{R}^T F - \id) &=
  \tr(\nabla\mathbf{u} - \overline{A}) + \text{h.o.t.} =
  \tr\nabla\mathbf{u} + \text{h.o.t.}
  \nonumber \\
  \sym(\overline{R}^T F - \id) &=
  \sym(\nabla\mathbf{u} - \overline{A}) + \text{h.o.t.} =
  \sym\nabla\mathbf{u} + \text{h.o.t.} \,,
\end{align}
since $\tr(\overline{A}) = 0$ and also $\sym(\overline{A}) = 0$. Thus $V_{\rm elastic}$ in the linearised approximation is given by
\begin{align}
  V_{\rm elastic} = \mu \,\| \sym \nabla\mathbf{u} \|^2 +
  \frac{\lambda}{2}\tr(\sym \nabla\mathbf{u})^2
  \label{Velas1}
\end{align}
which is the classical elastic energy of linear elasticity.

The term $\overline{R}^T\negthickspace \Curl \overline{R}$ simply becomes
\begin{align}
  \overline{R}^T\negthickspace \Curl \overline{R} =
  \Curl \overline{A} + \text{h.o.t.} \,.
\end{align}

Lastly, we need to consider $\polar(F)$. If $F = \id + \nabla\mathbf{u}$, then one verifies that $U^2 = F^T F = \id + \nabla\mathbf{u} + \nabla\mathbf{u}^T + \text{h.o.t.} = \id + 2 \sym \nabla\mathbf{u} + \text{h.o.t.}$ and also that $(\id + \sym \nabla\mathbf{u})^2 = \id + 2 \sym \nabla\mathbf{u} + \text{h.o.t.}$ so that we can write $U = \id + \sym \nabla\mathbf{u} + \text{h.o.t.}$ which yields the result
\begin{align}
  \polar(F) &= U^{-1} F = (\id - \sym \nabla\mathbf{u})(\id + \nabla\mathbf{u}) + \text{h.o.t.}
  \nonumber \\ &=
  (\id - \sym \nabla\mathbf{u})
  (\id + \sym \nabla\mathbf{u} + \skewp \nabla\mathbf{u}) + \text{h.o.t.}
  \nonumber \\ &=
  \id + \skewp \nabla\mathbf{u} + \text{h.o.t.} \,.
\end{align}
Therefore we can write the Cosserat coupling term in the completely linear approximation as follows
\begin{align}
  V_{\rm coupling} &=
  \mu_c\, \| \overline{R}^T\negthickspace \polar(F) - \id \|^2 =
  \mu_c\, \| (\id - \overline{A}) (\id + \skewp \nabla\mathbf{u}) - \id \|^2
  + \text{h.o.t.}
  \nonumber \\ &=
  \mu_c\, \| \id - \overline{A} + \skewp \nabla\mathbf{u} - \id \|^2
  + \text{h.o.t.}
  \nonumber \\& = \mu_c\, \| \skewp \nabla\mathbf{u} - \overline{A} \|^2
  + \text{h.o.t.} \,.
\end{align}

\subsection{A non-linear curvature model}

In the following we will consider a slightly different model which is different to the quadratic approximation as it is non-linear in the rotations. As before, we will assume that displacements are small $\|\mathbf{u}\| \ll 1$ and $\|\nabla\mathbf{u}\| \ll 1$, but we allow for arbitrary large  rotations $\overline{R}=\exp{\overline{A}}$. However, we will assume that the terms $\nabla\mathbf{u} \,\overline{A}^n$ are small, this means $\|\nabla\mathbf{u}\, \overline{A}^n\| \ll 1$ for all $n \geq 1$.

In this case, we again find
\begin{align}
  \overline{R}^T F - \id = \nabla\mathbf{u} - \overline{A} + \text{h.o.t.}
\end{align}
and our elastic energy is still~(\ref{Velas1}). The curvature energy~(\ref{Vcurv}) is unchanged and the interaction energy~(\ref{Vint}) is linearised in $\mathbf{u}$, as before. The Cosserat coupling term requires a closer look. Namely we find
\begin{align}
  V_{\rm coupling} &=
  \mu_c \,\| \overline{R}^T \polar(F) - \id \|^2
  \nonumber \\ &=
  \mu_c\, \| \overline{R}^T (\id + \skewp \nabla\mathbf{u}) - \id \|^2
  + \text{h.o.t.}
  \nonumber \\ &=
  \mu_c\, \| \overline{R}^T + \overline{R}^T \skewp \nabla\mathbf{u} - \id \|^2
  + \text{h.o.t.}
  \nonumber \\ &=
  \mu_c\, \| \overline{R}^T + (\id + \overline{A} + \overline{A}^2/2 + \ldots)^T \skewp \nabla\mathbf{u} - \id \|^2
  + \text{h.o.t.}
  \label{Vcoup_h1} \\ &=
  \mu_c \,\| \overline{R}^T - \id + \skewp \nabla\mathbf{u} \|^2
  + \text{h.o.t.}
  \nonumber \\ &=
  \mu_c\, \| \skewp \nabla\mathbf{u} - (\overline{R} - \id)\|^2
  + \text{h.o.t.} \,.
\end{align}
The crucial step is in~(\ref{Vcoup_h1}) where we used the definition of the matrix exponential and our assumption that $\|\nabla\mathbf{u} \overline{A}^n\| \ll 1$ but did not assume the rotations to be small. Combining this into a single energy functional gives us our non-linear curvature model
\begin{align}
  V_2 = V_{\rm elastic} + V_{\rm curvature} + V_{\rm interaction} + V_{\rm coupling},
\end{align}
where the individual parts are given by
\begin{align*}
  V_{\rm elastic} = {} &\mu\, \| \sym \nabla\mathbf{u} \|^2 +
  \frac{\lambda}{2}\,\tr(\sym \nabla\mathbf{u})^2 \\
  V_{\rm curvature} =
  {} &\kappa_1 \,\| \dev \sym (\overline{R}^T\negthickspace \Curl \overline{R})\|^2 +
  \kappa_2 \,\| \skewp (\overline{R}^T\negthickspace \Curl \overline{R})\|^2
  \nonumber \\& +
  \kappa_3 \,\tr(\overline{R}^T\negthickspace \Curl \overline{R})^2 \\
  V_{\rm interaction} =
  {} &\chi_1\, \tr(\overline{R}^T\negthickspace \Curl \overline{R}) \tr(\nabla\mathbf{u})
  \nonumber \\&+
  \chi_3\, \langle \dev \sym (\overline{R}^T\negthickspace \Curl \overline{R}),
  \dev \sym \nabla\mathbf{u}\rangle \\
  V_{\rm coupling} = {} &\mu_c\, \| \skewp \nabla\mathbf{u} - (\overline{R} - \id) \|^2.
\end{align*}

The equations of motion are derived using the calculus of variations. We will vary the Lagrangian $V_2$ with respect to $\delta u_i$ and $\delta \overline{A}_{ij}$. However, it is more convenient to work with the vector $\mathbf{a} = (a_1,a_2,a_3)$ such that $\overline{A}_{ij} = \varepsilon_{ikj} a_k$ and consider the variations with respect to this vector, we note $\delta \overline{A}_{ij} = \varepsilon_{ikj} \delta a_k$. We use the notation $a = \sqrt{a_1^2+a_2^2+a_3^2}$ for the norm of the vector $\mathbf{a}$. The variations with respect to $\delta u_i$ are pretty straightforward to calculate. This cannot be said for the variations with respect to $\delta a_k$. We note that the variations of the rotation matrix $\overline{R}$ with respect to $\delta a_k$ are already very complicated, the result is stated for convenience
\begin{align}
  \frac{\delta R_{ij}}{\delta a_k} =
  {} &\left(\frac{a \cos a-\sin a}{a^3} \right) a_k a_l \,\varepsilon_{ilj}+\frac{\sin a}{a}\varepsilon_{ikj}
  \nonumber \\ &+
  {} \frac{2(\cos a-1)+a \sin a}{a^4} a_k(a_i a_j-a_n a_n \delta_{ij})
  \nonumber \\ &+
  {} \left(\frac{1-\cos a}{a^2}\right)(a_i\delta_{jk}+a_j\delta_{ik}-2a_k\delta_{ij}).
\end{align}
It becomes clear that the variations of $V_2$ with respect to $\delta a_k$ will result in very long expressions which we will not state explicitly. In particular, the variations of $V_{\rm curvature}$ will result in complicated expressions since this potential is fourth order in $\overline{R}$.

Let us now consider the variations of the functional $V_2$ with respect to the displacements $u_i$. This results in the following equation using the index notation
\begin{align}
  \rho\, \partial_{tt} u_i = {} &(\mu+\mu_c)\,u_{i,jj} + (\mu+\lambda-\mu_c)\,u_{j,ji}
  \nonumber \\ &
  + \frac{1}{3}(3\chi_1-\chi_3)\,(R^T \Curl R)_{jj,i}
  + \chi_3\, (\sym (R^T \Curl R))_{ij,j}
  \nonumber \\ &
  + \mu_c (R_{ji,j}-R_{ij,j}).
\end{align}
We can also write this in matrix notation. Using the matrix divergence, we have
\begin{align}
  \rho\, \partial_{tt} \mathbf{u} = {} &(\mu+\mu_c)\Delta \mathbf{u} + (\mu+\lambda-\mu_c)\grad \divv \mathbf{u}
  \nonumber \\ &
  + \chi_1 \grad \tr(R^T \Curl R)
  + \chi_3 \Divv [\dev \sym (R^T \Curl R)]
  \nonumber \\ &
  - 2 \mu_c \Divv [\skewp \overline{R}] .
  \label{eom_matform}
\end{align}
This is the equation of motion for the displacement $\mathbf{u}$. The corresponding equations of motion for the rotations are considerably longer.

Alternatively, we can write~(\ref{eom_matform}) using a slightly different notation involving the matrix $\nabla \mathbf{u}$. We have the following identities 
\begin{align}
  \mu\, \Delta \mathbf{u} + (\mu+\lambda) \grad \divv \mathbf{u} &= 
  \Divv \left[2 \mu \sym \nabla \mathbf{u} + \lambda \tr(\nabla \mathbf{u}) \id \right], \\
  \mu_c\, \Delta \mathbf{u} - \mu_c \grad \divv \mathbf{u} &=
  2 \mu_c \Divv [\skewp \nabla \mathbf{u}].
\end{align}
Therefore we can write~(\ref{eom_matform}) as follows
\begin{align}
  \rho\, \partial_{tt} \mathbf{u} = 
  {} &\Divv \left[2 \mu \sym \nabla \mathbf{u} + \lambda \tr(\nabla \mathbf{u}) \id \right]
  \nonumber \\
  &+ \chi_1 \grad \tr(R^T \Curl R) + \chi_3 \Divv [\dev \sym (R^T \Curl R)] 
  \nonumber \\
  &+ 2 \mu_c \Divv [\skewp \nabla \mathbf{u}] - 2 \mu_c \Divv [\skewp \overline{R}].
  \end{align}
Lastly, using that $\skewp (\overline{R} - \id)$ = $\skewp \overline{R}$ we arrive at
\begin{align}
  \rho\, \partial_{tt} \mathbf{u} = 
  {} &\Divv \left[
  2 \mu \sym \nabla \mathbf{u} + \lambda \tr(\nabla \mathbf{u}) \id 
  + 2 \mu_c \skewp (\nabla \mathbf{u} - (\overline{R}-\id))
  \right]
  \nonumber \\
  &+ \chi_1 \grad \tr(R^T \Curl R) + \chi_3 \Divv [\dev \sym (R^T \Curl R)],
  \end{align}
which is a neat form of the equation for the displacements.

\section{Solitonic solutions}

\subsection{Assumptions}

The full equations of motion for the rotations cannot be explicitly written down since they would be too long. We can simplify our equations considerably by making several assumptions, most of which are related to the ones used in~\cite{boehmer2013rota}.

First of all, let us assume that the points in our continuum can only experience rotations about one axis, the $z$-axis say, which means we choose
\begin{align}
  \overline{A} =
  \begin{pmatrix}
    0 & -\phi &0\\
    \phi & 0 &0\\
    0 & 0 &0\\
  \end{pmatrix},
  \qquad
  \overline{R} = e^{\overline{A}} =
  \begin{pmatrix}
    \cos \phi & -\sin \phi & 0\\
    \sin \phi & \cos \phi & 0\\
    0 & 0 & 1\\
  \end{pmatrix},
\end{align}
where $\phi=\phi(x,y,z,t)$ is an arbitrary function of the three spatial variables and time. Our choice of the rotation axis sets both $a_1$ and $a_2$ to zero. Analogously, we assume that the displacements must occur along the axis of rotation. Therefore, we have that $u_1$ and $u_2$ vanish, and $u_3=\psi(x,y,z,t)$, with $\psi$ also being an arbitrary function of the three spatial variables and time.

It is well known that there are two possible types of plane waves in classical elasticity: transverse and longitudinal waves. Since we assumed that our wave propagates along the $z$-axis, we see that if the elastic wave is longitudinal, we have that $\psi=\psi(z,t)$. On the other hand, if the elastic wave is transverse, then $\psi$ should only depend on the coordinates orthogonal to the $z$-axis, i.e.~$\psi=\psi(x,y,t)$. While it is less clear what it means for a rotational wave to be transverse or longitudinal, we will simply use the above conventions. For a longitudinal wave we would write $\phi=\phi(z,t)$ while for a transverse rotational wave we assume $\phi=\phi(x,y,t)$. Since we are interested in finding wave type solutions of the equations of motions, we can in principle study four different combinations of the above waves.

In the following we will discuss the equations and their solutions assuming that both, the elastic and the rotational wave are both longitudinal waves, this means we assume $\psi=\psi(z,t)$ and $\phi=\phi(z,t)$, or expressed in our original variables
\begin{align}
  \mathbf{u} =
  \begin{pmatrix}
    0 \\ 0 \\ \psi(z,t)
  \end{pmatrix},
  \qquad
  F =
  \begin{pmatrix}
    1 & 0 & 0\\
    0 & 1 & 0\\
    0 & 0 & 1 + \partial_{z} \psi\\
  \end{pmatrix},
\end{align}
for the desplacements. For the rotations we have
\begin{align}
  \overline{R} =
  \begin{pmatrix}
    \cos \phi(z,t) & -\sin \phi(z,t) & 0\\
    \sin \phi(z,t) & \cos \phi(z,t) & 0\\
    0 & 0 & 1\\
  \end{pmatrix},
  \qquad
  \overline{R}^T\negthickspace \Curl \overline{R} =
  \begin{pmatrix}
    \partial_{z} \phi & 0 & 0\\
    0 & \partial_{z} \phi & 0\\
    0 & 0 & 0\\
  \end{pmatrix}.
\end{align}
This ansatz for the solution can be interpreted in the following way. Along the z-axis we have a longitudinal wave which we can view as a compression wave. This axis also defines our axis for rotations which means we consider rotational waves transversal to the compression wave.

\subsection{The equations of motion}

Using our ansatz, the six independent equations of motion reduce to two equations, one for the field $\phi$ and one for the field $\psi$. They are given by
\begin{align}
  \frac{1}{3}(\kappa_1+6\kappa_3) \partial_{zz}\phi +
  \frac{1}{6}(3\chi_1-\chi_3)\partial_{zz}\psi -
  \mu_\mathrm{c} \sin(\phi) -
  \rho_{\mathrm{rot}} \partial_{tt}\phi &= 0,
  \label{eq:rot} \\
  (\lambda+2\mu)\partial_{zz}\psi +
  \frac{2}{3}(3\chi_1-\chi_3)\partial_{zz}\phi-\rho\, \partial_{tt}\psi &= 0.
  \label{eq:elastic}
\end{align}

The two equations can be rewritten, using matrix notation, as follows:
\begin{align}
  \begin{pmatrix}
    \partial_{tt}\phi \\
    \partial_{tt}\psi
  \end{pmatrix}	=
  \mathbf{M}
  \begin{pmatrix}
    \partial_{zz}\phi \\
    \partial_{zz}\psi
  \end{pmatrix}-
  \frac{\mu_\mathrm{c}}{\rho_{\mathrm{rot}}}
  \begin{pmatrix}
    \sin(\phi)\\
    0
  \end{pmatrix},
  \label{eq:mform}
\end{align}
where the matrix $\mathbf{M}$ is given by
\begin{align}
  \mathbf{M} =
  \begin{pmatrix}
    (\kappa_1+6\kappa_3)/(3\rho_{\mathrm{rot}}) &
    (3\chi_1-\chi_3)/(6\rho_{\mathrm{rot}}) \\
    2(3\chi_1-\chi_3)/(3\rho) &
    (\lambda+2\mu)/\rho
  \end{pmatrix}.
\end{align}
We note that the component $M_{22}$ is the square of the elastic (longitudinal) wave speed and we will use the notation $M_{22} = v_{\rm elas}^2$. Similarly, the component $M_{11}$ corresponds to the square of the rotational wave speed and we will write $M_{11} = v_{\rm rot}^2$. It turns out to be convenient to work with the components of $\mathbf{M}$ instead of the elastic moduli of the energy functional.

The system of equations (\ref{eq:rot})--(\ref{eq:elastic}) shows similarities with the sine-Gordon-d'Alembert system, see for instance~\cite{Kivshar_Malomed:1989}, which has been previously studied in the context of micropolar elasticity and gives rise to soliton-like solutions~\cite{Maugin_Miled:1986a,Maugin_Miled:1986b,Puget_Maugin:1989,Sayadi_Puget:1991}. The key difference between our system and those studied previously is the presence of interaction terms due to the interaction Lagrangian~(\ref{Vint}) which results in additional second derivative terms. Moreover, the sine-Gordon-d'Alembert system typically contains first derivative terms of the field variables which are also absent is our approach. It is likely that a more general ansatz will result in equations of motion which show a greater similarity to the sine-Gordon-d'Alembert system, however, the interaction terms due to $V_{\rm interaction}$ will make this system uniquely different again. 

\subsection{Sine-Gordon type equation}

We start by recalling the scalar wave equation $\partial_{tt}\varphi(t,x) - c^2 \partial_{xx}\varphi(t,x)= 0$ for the function $\varphi(t,x)$ where $c$ is the wave speed. A very important and related equation is the sine-Gordon equation $\partial_{tt} \varphi(t,x) - \partial_{xx}\varphi(t,x) + \sin(\varphi(t,x)) = 0$ which is known to have soliton solutions.

Inspection of~(\ref{eq:mform}) motivates us to seek soliton type solutions of this coupled system of equations. Equation~(\ref{eq:rot}) contains the term $(3\chi_1-\chi_3)/(6\rho_{\mathrm{rot}}) \partial_{zz} \psi$. If we can `remove' this term we would obtain a sine-Gordon equation for $\phi$. The easiest way to obtain a sine-Gordon equation would be to simply set $\chi_1 = \chi_3 = 0$ in which case the displacement $u_3 = \psi$ will satisfy the wave equation, while the rotation $\phi$ will be of the form of a sine-Gordon equation. This rotational sine-Gordon equation was discussed in~\cite{boehmer2012gauge}.

In the following we will demonstrate how this can be achieved. Start with a plane wave ansatz for $\psi$ in the form
\begin{align}
  \psi = g(z-v_1 t),
  \label{eq:travel}
\end{align}
which corresponds to a travelling wave in the positive $z$ direction only, and $g$ is an arbitrary function of its argument, and $v_1$ is a wave speed. Due to the complicated structure of the equations, we were not able to consider the more general wave ansatz $f(z+v_1t) + g(z-v_1t)$. Now that $\psi$ satisfies the wave equation $\partial_{tt} \psi = v_1^2\, \partial_{zz} \psi$, this is substituted into~(\ref{eq:elastic}) to obtain
\begin{align}
  g''(z-v_1t) = \frac{M_{21}}{v_1^2-v_{\rm elas}^2}\partial_{zz} \phi,
  \label{gpp}
\end{align}
where we assume that $v_1^2-a_{22} \neq 0$. Next, relation~(\ref{gpp}) is used so that $\partial_{zz} \psi$ can be eliminated from~(\ref{eq:rot}), we thus arrive at
\begin{align}
  \partial_{tt}\phi -
  \left(v_{\rm rot}^2+\frac{M_{12}\,M_{21}}{v_1^2-v_{\rm elas}^2} \right)
  \partial_{zz} \phi
  + \frac{\mu_\mathrm{c}}{\rho_{\mathrm{rot}}} \sin\phi = 0,
\end{align}
which is the sine-Gordon equation. It could be brought into the above mentioned standard form by rescaling the spatial coordinate $z$ as follows
\begin{align}
  z = \left(v_{\rm rot}^2+\frac{M_{12}\,M_{21}}{v_1^2-v_{\rm elas}^2} \right)^{1/2}
  \hat{z}
\end{align}
and introducing the parameter $m^2 = \mu_\mathrm{c}/\rho_{\mathrm{rot}}$, and we would arrive at
\begin{align}
  \partial_{tt}\phi - \partial_{\hat{z}\hat{z}} \phi
  + m^2 \sin\phi = 0.
\end{align}
However, in the following we will work with our original variables.

\subsection{Solitonic solutions}

We are now seeking solitonic solutions for the function $\phi$ and assume it be of the form	
\begin{align}
  \phi = 4 \arctan e^{k (z-v_2 t)+\delta},
  \label{eq:soliton}
\end{align}
where $v_2$, $k$ and $\delta$ are constants. This sine-Gordon ansatz is substituted into Eqs.~(\ref{gpp}) with the aim of solving for $g(z-v_1t)$. In order to proceed we must make one additional assumption, namely that the elastic and the rotational wave propagate with the same wave speed $v_1 = v_2 = v$. By substituting the ansatz~(\ref{eq:soliton}) along with $s=z-v\,t$ into~(\ref{eq:elastic}) we arrive at the following differential equation for $g(s)$
\begin{align}
  g''(s)=\frac{4 M_{21}\, k^2}{v_{\rm elas}^2-v^2}\,
  \frac{ e^{k s+\delta} (e^{2(k s+\delta)}-1)}{(e^{2(k s+\delta)}+1)^2},
\end{align}
which can be integrated twice with respect to $s$ to give
\begin{align}
  g(s)=-\frac{4 M_{21}}{v_{\rm elas}^2-v^2}
  \arctan e^{k s+\delta} + C_1 + C_2 s,
  \label{eq:g}
\end{align}
where $C_1$ and $C_2$ are two constants of integration which we set to zero.

Next, we need to substitute the ansatz~(\ref{eq:soliton}) with~(\ref{eq:g}) into the remaining equation~(\ref{eq:rot}). This yields an algebraic equation which relates the various constants of the problem. We find, after some algebra,
\begin{align}
  k^2 (v^4-\tr(M)v^2+\det(M)) - (v_{\rm elas}^2-v^2)
  \frac{\mu_c}{\rho_{\mathrm{rot}}} = 0,
  \label{eq:condition}
\end{align}
which fixes either the parameter $k$ or the velocity $v$, and we should note that $\tr(M) = v_{\rm elas}^2 + v_{\rm rot}^2$. The term $\det(M)$ also depends on the coupling constants. This relation for $k$ is given by
\begin{align}
  k^2 = \frac{v_{\rm elas}^2-v^2}{v^4-\tr(M)v^2+\det(M)}
  \frac{\mu_c}{\rho_{\mathrm{rot}}}.
  \label{eq:k}
\end{align}
The solution with positive root $k_{+}$ is called a kink, while the solution with the negative root $k_{-}$ is known as the antikink, see~\cite{drazin1989soliton}.

We can also use equation~(\ref{eq:condition}) to find $v^2$ in terms of $k$, $m$ and each $M_{ij}$. As expected, letting $\mu_c=0$ in~(\ref{eq:condition}) gives us an equation which determines the velocities independent of $k$. The resulting velocities are those reported in equation (4.18) of~\cite{boehmer2013rota}, since this is equivalent to setting $\mu_{\mathrm{c}}=0$.

Putting everything together, our solution is given explicitly by
\begin{align}
  \phi &= 4 \arctan e^{k_{\pm} (z-v t)+\delta}, \\
  \psi &= - \frac{M_{21}}{v_{\rm elas}^2-v^2} 4 \arctan e^{k_{\pm} (z-v t)+\delta},
\end{align}
with $k_{\pm}$ given by~(\ref{eq:k}) and $\delta$ and $v$ being constants.

\begin{figure}[htb!]
  \centering
  \includegraphics[width=0.7\textwidth]{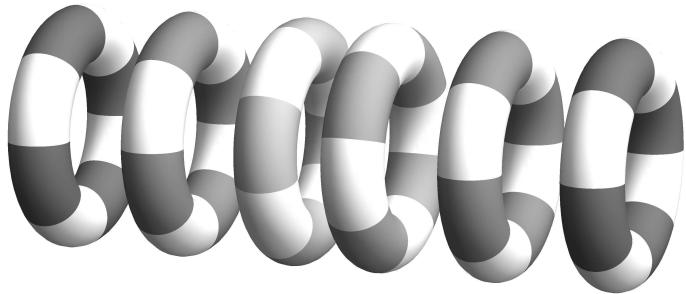}
  \caption{Visualisation of the soliton solution. The material points are represented by tori, the microrotations are visualised by the rotations of the tori.}
  \label{fig:soliton_f15}
\end{figure}

We visualise this solution in Fig.~\ref{fig:soliton_f15}. The $z$-axis is along the horizontal. Every material point with microstructure is represented by a ring (torus). The grey pattern on the rings helps to visualise the rotations about the $z$-axis. Looking closely at the figure, we see that rings 1 \& 2, and 5 \& 6 are in their (almost) undeformed states while the positions of rings 3 \& 4 are displaced. Firstly, we note that the spatial separation between the middle rings is smaller than the one of the outer rings. This displacement is due to the longitudinal elastic soliton travelling along that axis. Secondly, we note that the middle two rings are rotated with respect to the outer rings. This visualises the microrotational soliton of the microstructure.

\section{Discussion}

The geometrically nonlinear Cosserat micropolar elasticity model with nonstandard constitutive coupling between Cosserat rotations and the deformation gradient was studied. We were able to find soliton type solutions assuming small displacement but allowing arbitrarily large Cosserat rotations. In general the Euler-Lagrange equations of this system are very complicated, however, using a specific ansatz, it is possible to significantly simplify these equations. We assume a compression wave along the $z$-axis and restrict possible rotations to be about this axis. The corresponding solution we find consists of two solitonic waves, an elastic one and the microrotational one, both of which have the same wave speed. 

The existence of such soliton type solutions in our model is related to the Cosserat couple term. In the limit $\mu_c \rightarrow 0$ our equations of motion would not have such a solution, and we would instead find two coupled plane waves. It is precisely this Cosserat couple modulus $\mu_c$ which appears to localise these waves. At this stage it is not clear whether this statement can be generalised to wave type solutions of different form, our ansatz is very specific. However, these results point into a possible direction of interpreting the Cosserat couple modulus. 

The dynamical Cosserat model studied here is particularly interesting from various points of view. The model is intrinsically nonlinear because rotations in three dimensions do not commute, in other words the group $\mathrm{SO}(3)$ is non-Abelian. In the case of planar Cosserat elasticity, the matrix rotation $\overline{R}$ has one degree of freedom only, one can only rotate about one axis, and these rotations commute. The group $\mathrm{SO}(2)$ is Abelian. Hence, one cannot meaningfully simplify the dynamical Cosserat model to the planar case as essential features of the model would be lost. It will be important to study the complete model further, clearly the next significant task is to study the fully nonlinear model where the displacements are not assumed to be small. This means studying a continuum with arbitrarily large displacements and arbitrarily large rotations. One should think that a suitable ansatz will simplify these equations to the point where one can seek explicit solutions. 


\begin{thebibliography}{10}


\bibitem{boehmer2011rota}
C.~G. B{\"{o}}hmer, R.~J. Downes, and D.~Vassiliev.
\newblock Rotational elasticity.
\newblock {\em Q. J. Mechanics Appl. Math.}, 64(4):415--439, 2011.

\bibitem{boehmer2012gauge}
C.~G. B{\"{o}}hmer and Y.~N. Obukhov.
\newblock A gauge-theoretic approach to elasticity with microrotations.
\newblock {\em Proc. R. Soc. A}, 468:1391--1407, 2012.

\bibitem{boehmer2013rota}
C.~G. B{\"{o}}hmer and N.~Tamanini.
\newblock Rotational elasticity and couplings to linear elasticity.
\newblock {\em to appear in Math. Mech. Solids}, 2013.

\bibitem{Cosserat09}
E.~Cosserat and F.~Cosserat.
\newblock {\em Th\'eorie des corps d\'eformables.}
\newblock Librairie Scientifique A. Hermann et Fils (engl. translation by D.
  Delphenich 2007, pdf available at
  http://www.mathematik.tu-darmstadt.de/fbereiche/analysis/pde/staff/neff/patrizio/{C}osserat.html),
  reprint 2009 by Hermann Librairie Scientifique, ISBN 978 27056 6920 1, Paris,
  1909.

\bibitem{drazin1989soliton}
P.~G. Drazin and R.~S. Johnson.
\newblock {\em Solitons: An Introduction}.
\newblock Cambridge University Press, 1989.





\bibitem{Ericksen:1962b}
J.~L. Ericksen.
\newblock Hydrostatic theory of liquid crystals.
\newblock {\em Arch. Rational Mech. Anal.}, 9:379--394, 1962.

\bibitem{Ericksen:1962a}
J.~L. Ericksen.
\newblock Kinematics of macromolecules.
\newblock {\em Arch. Rational Mech. Anal.}, 9:1--8, 1962.

\bibitem{Ericksen:1967}
J.~L. Ericksen.
\newblock Twisting of liquid crystals.
\newblock {\em J. Fluid Mech.}, 27:59--64, 1967.

\bibitem{Ericksen:1957}
J.~L. Ericksen and C.~Truesdell.
\newblock Exact theory of stress and strain in rods and shells.
\newblock {\em Arch. Rational Mech. Anal.}, 1:295--323, 1957.

\bibitem{Eringen:1999}
A.~C. Eringen.
\newblock {\em Microcontinuum field theories: I. Foundations and solids}.
\newblock Springer, New York, 1999.

\bibitem{Eringen:1964a}
A.~C. Eringen and E.~S. Suhubi.
\newblock Nonlinear theory of simple microelastic solids {I}.
\newblock {\em Int. J. Eng. Sci.}, 2:189--204, 1964.

\bibitem{Eringen:1964b}
A.~C. Eringen and E.~S. Suhubi.
\newblock Nonlinear theory of simple microelastic solids {II}.
\newblock {\em Int. J. Eng. Sci.}, 2:389--404, 1964.

\bibitem{Green:1964}
A.~E. Green.
\newblock Multipolar continuum mechanics.
\newblock {\em Arch. Rational Mech. Anal.}, 17:113--147, 1964.





\bibitem{Kivshar_Malomed:1989}
Y.~S. Kivshar and B.~A. Malomed.
\newblock Dynamics of solitons in nearly integrable systems.
\newblock {\em Rev. Mod. Phys.}, 61:763--915, Oct 1989.

\bibitem{Kivshar_Malomed:1990}
Y.~S. Kivshar and B.~A. Malomed.
\newblock Dynamics of domain walls in elastic ferromagnets and ferroelectrics.
\newblock {\em Phys. Rev.}, B42:8561--8570, 1990.

\bibitem{kroener}
E.~Kr{\"o}ner.
\newblock {\em Mechanics of generalized continua}.
\newblock Springer, 1967.
\newblock Proc. IUTAM Symp. Freudenstadt-Stuttgart.


\bibitem{Neff_Ghiba_band_gaps_micromorph13}
A.~Madeo, P.~Neff, I.~D. Ghiba, L.~Placidi, and G.~Rosi.
\newblock Wave propagation in relaxed micromorphic continua: modeling
  metamaterials with frequency band-gaps.
\newblock {\em Cont. Mech. Thermod.}, 2013.

\bibitem{Neff_Ghiba_band_gaps_micromorph_zamm14}
A.~Madeo, P.~Neff, I.~D. Ghiba, L.~Placidi, and G.~Rosi.
\newblock Band gaps in the relaxed linear micromorphic continuum.
\newblock {\em to appear in Z. Angew. Math. Mech.}, 2014.

\bibitem{Maugin_Miled:1986b}
G.~A. Maugin and A.~Miled.
\newblock Solitary waves in elastic ferromagnets.
\newblock {\em Phys. Rev.}, B33(7):4830--4842, 1986.

\bibitem{Maugin_Miled:1986a}
G.~A. Maugin and A.~Miled.
\newblock Solitary waves in micropolar elastic crystals.
\newblock {\em Int. J. Engng. Sci.}, 24(9):1477--1499, 1986.

\bibitem{Mindlin1964}
R.~D. Mindlin.
\newblock Micro-structure in linear elasticity.
\newblock {\em Arch. Rational Mech. Anal.}, 16(1):51--78, 1964.


\bibitem{Neff_micromorphic_rse_05}
P.~Neff.
\newblock Existence of minimizers for a finite-strain micromorphic elastic
  solid.
\newblock {\em Proc. Roy. Soc. Edinb. A}, 136:997--1012, 2006.

\bibitem{Neff_Cosserat_plasticity05}
P.~Neff.
\newblock A finite-strain elastic-plastic {C}osserat theory for polycrystals
  with grain rotations.
\newblock {\em Int. J. Eng. Sci.}, 44:574--594, 2006.

\bibitem{Neffexistence:2014}
P.~Neff, M.~Birsan, and F.~Osterbrink.
\newblock Existence theorem for geometrically nonlinear {C}osserat micropolar
  model under uniform convexity requirements.
\newblock {\em to appear in J. Elasticity}, 2015.

\bibitem{Neff_Forest_jel05}
P.~Neff and S.~Forest.
\newblock A geometrically exact micromorphic model for elastic metallic foams
  accounting for affine microstructure. {M}odelling, existence of minimizers,
  identification of moduli and computational results.
\newblock {\em J. Elasticity}, 87:239--276, 2007.



\bibitem{Neff_Ghiba_unified_micromorph13}
P.~Neff, I.D. Ghiba, A.~Madeo, L.~Placidi, and G.~Rosi.
\newblock A unifying perspective: the relaxed linear micromorphic continuum.
\newblock {\em Cont. Mech. Thermodyn.}, 26:639--681, 2014.


\bibitem{Neff_curl06}
P.~Neff and I.~M\"unch.
\newblock Curl bounds {Grad} on {${\rm SO}(3)$}.
\newblock {\em ESAIM: Control, Optimisation and Calculus of Variations},
  14(1):148--159, 2008.

\bibitem{Neff_Muench_simple_shear09}
P.~Neff and I.~M\"unch.
\newblock Simple shear in nonlinear {Cosserat} elasticity: bifurcation and
  induced microstructure.
\newblock {\em Cont. Mech. Thermod.}, 21(3):195--221, 2009.


\bibitem{Potapov:1999}
A.~I. Potapov, I.~S. Pavlov, and G.~A. Maugin.
\newblock Nonlinear wave interactions in {1D} crystals with complex lattice.
\newblock {\em Wave Motion}, 29(4):297 -- 312, 1999.

\bibitem{Puget_Maugin:1989}
J.~Pouget and G.~A. Maugin.
\newblock Nonlinear dynamics of oriented elastic solids -- {II}. propagation of
  solitons.
\newblock {\em J. of Elasticity}, 22:157--183, 1989.






\bibitem{Sayadi_Puget:1991}
M.~K. Sayadi and J.~Pouget.
\newblock Soliton dynamics in a microstructured lattice model.
\newblock {\em J. Physics A: Gen. Phys.}, 24:2151--2172, 1991.

\bibitem{Schaefer:1967}
H.~Schaefer.
\newblock Das {C}osserat {K}ontinuum.
\newblock {\em Z. Angew. Math. Mech.}, 47:485--498, 1967.

\bibitem{Toupin:1962}
R.~A. Toupin.
\newblock Elastic materials with couple-stresses.
\newblock {\em Arch. Rational Mech. Anal.}, 11:385--414, 1962.

\bibitem{Toupin:1964}
R.~A. Toupin.
\newblock Theories of elasticity with couple-stress.
\newblock {\em Arch. Rational Mech. Anal.}, 17:85--112, 1964.

\bibitem{Whittaker_VolI}
E.~Whittaker.
\newblock {\em A History of the Theories of Aether and Electricity}.
\newblock Thomas Nelson and Sons, 1951.

\bibitem{Yavari:2012}
A.~Yavari and A.~Goriely.
\newblock {R}iemann-{C}artan geometry of nonlinear dislocation mechanics.
\newblock {\em Arch. Rational Mech. Anal.}, 205:59--118, 2012.

\bibitem{Yavari:2013}
A.~Yavari and A.~Goriely.
\newblock {R}iemann-{C}artan geometry of nonlinear disclination mechanics.
\newblock {\em Math. Mech. Solids}, 18:91--102, 2013.

\bibitem{Zorski_Infeld:1992}
H.~Zorski and E.~Infeld.
\newblock New soliton equation for dipole chains.
\newblock {\em Phys. Rev. Lett.}, 68:1180--1183, 1992.

\end{thebibliography}

\end{document}